\documentstyle [12pt]{article}

\newcommand{\be}{\begin{equation}}
\newcommand{\ee}{\end{equation}}

\catcode`\@=11       %
\def\uti#1{\mathop{\vtop{\ialign{##\crcr
      $\hfil\displaystyle{#1}\hfil$\crcr\noalign{\kern3\p@\nointerlineskip}
      $\scriptstyle\sim$\crcr\noalign{\kern3\p@}}}}\limits}
\catcode`\@=12

\catcode`\@=11       %
\def\mar#1{\mathop{\vtop{\ialign{##\crcr
      $\hfil\displaystyle{#1}\hfil$\crcr\noalign{\kern3\p@\nointerlineskip}
      $M\rightarrow\infty$\crcr\noalign{\kern3\p@}}}}\limits}
\catcode`\@=12

\catcode`\@=11       %
\def\mxr#1{\mathop{\vtop{\ialign{##\crcr
      $\hfil\displaystyle{#1}\hfil$\crcr\noalign{\kern3\p@\nointerlineskip}
      $M\to \infty\choose x\to -\infty$\crcr\noalign{\kern3\p@}}}}\limits}
\catcode`\@=12

\def\undersymbolsub#1#2{\mathop{\vtop{\ialign{##\crcr
$\hfil\displaystyle{#2}\hfil$\crcr\noalign
{\kern1pt\nointerlineskip}\hbox{$\hfil\scriptstyle{#1}\hfil$}\crcr
\noalign{\kern1pt}}}}}

\def\undersymbolsub#1#2{\mathop{\vtop{\ialign{##\crcr
$\hfil\displaystyle{#2}\hfil$\crcr\noalign
{\kern1pt\nointerlineskip}\hbox{$\hfil\scriptstyle{#1}\hfil$}\crcr
\noalign{\kern1pt}}}}}

\def\overlay#1#2{\setbox0=\hbox{#1}\setbox1=\hbox to \wd0{\hss #2\hss}#1%
\hskip -2\wd0\copy1}
\begin{document}

{\centerline {\bf {Searching a doubly charged Higgs boson at Hera }}}


\vskip 1 truecm
\centerline {Elena Accomando}
\centerline{\it  Dipartimento di Fisica, Universit\`a "La Sapienza"}
\centerline{\it P.le A. Moro 2, 00185 Roma, Italy}

\centerline {and}

\centerline {Silvano Petrarca}
\centerline{\it  Dipartimento di Fisica, Universit\`a di Roma "La Sapienza"}
\centerline{\it and INFN, Sezione di Roma "La Sapienza"}
\centerline{\it P.le A. Moro 2, 00185 Roma, Italy}
\vskip 2.5 truecm

\noindent ABSTRACT
\vskip 0.5 truecm
The production of a single exotic Higgs particle is studied at Hera. Within
the present limits on the Yukawa couplings this doubly charged particle,
suggested by the left-right symmetric models, can be observed at Hera up to
values of its mass of about 150 GeV.
\vskip .1 in
PACS numbers:

\newpage

The doubly charged Higgs bosons are the basic particles of a class of
models of electroweak interactions,
beyond the Standard Model, with spontaneous
parity violation \cite{SM},\cite{Gelmini}.
In the left-right symmetric model of Senjanovic and Mohapatra \cite{SM}, based
on the gauge group $G=SU(2)_L\otimes SU(2)_R\otimes U(1)_{B-L}$ \cite{SM},
the scalar sector contains
two triplets of Higgs fields $H_L$ and $H_R$ which
transform as (3,1,2) and (1,3,2) under $G$.

The phenomenology of the Higgs particle is
uniquely determined once its mass $M_H$ and its Yukawa couplings to the leptons
$g_{ll'}$ are given. Therefore, we briefly summarize the present experimental
limits on the above two parameters.
In the framework of the left-right symmetric model, by taking into account
the experimental limits on the neutrinoless double-$\beta$ decay of $Ge^{76}$,
Mohapatra found that $g_{ee}^2\le 0.1$ \cite{moha}. This bound
is in agreement with the
allowed values of $g_{ee}$ given in ref.\cite{Pic}, where the limits on the
$(\beta\beta )_{o\nu}$ decay of the $Te^{130}$ are also considered.

The case of a nondiagonal coupling (in the lepton flavour) of the $H^{--}$
to the charged leptons has been considered by Swartz \cite{Swartz}.
In his analysis, taking into account the spontaneous conversion of
muonium into antimuonium, the high-energy Bhabha scattering and the flavour
changing $\mu \rightarrow ee \overline{e} $ decay,
the following limits are established:
$g_{ee}g_{\mu\mu}<5.8\times 10^{-5}(M_H/1~GeV)^{2}$,
$g^2_{ee}<9.7\times 10^{-6}(M_H/1~GeV)^{2}$ as well as the most stringent one
\be
{g_{ee}g_{e\mu}\over{M_H^2}}<4.8\times 10^{-11}~GeV^{-2}
\label{1}
\ee

Concerning the values of the doubly charged Higgs masses,
the range $6.5~GeV\le M_H\le 36.5~GeV$
has been excluded by a recent experimental search of Higgs
triplets at the SLAC collider \cite{SLAC}. These bounds are obtained
assuming $g_{ll'}\ge 3\cdot 10^{-7}$ and examining the decay of the $Z$ boson
into a ($H^{++},H^{--}$) pair.

In ref.\cite{LEP} it has been pointed out that at LEP1 these masses
can be probed up to $M_H\sim 100~$GeV through the single $H^{--}$
production, when two alternative scenarios are assumed for the Yukawa
couplings: $g_{ee}\gg g_{e\mu}$ or $g_{e\mu}\gg g_{ee}$.

In this letter we adopt a point of view analogous to that of ref.\cite{LEP}
to the aim of discussing the possible production of a single doubly
charged Higgs boson at HERA, the new $ep$ collider in Hamburg, where
\cite{HERA} an electron beam with $E_e=30~GeV$ and a proton beam with
$E_p=820~GeV$ collide head-on ($\sqrt{s}\simeq 313~GeV$), with a
luminosity up to $1.5\cdot 10^{31}~cm^{-2}s^{-1}$, according to the project.

We calculated the following processes:
\be
e^-p\rightarrow e^+pH^{--}~~~~~~~~~~~~~~H^{--}\rightarrow e^-e^-(\mu^-\mu^-
,\tau^-\tau^-)
\label{2}
\end{equation}
\be
e^-p\rightarrow\mu^+pH^{--}~~~~~~~~~~~~~H^{--}\rightarrow e^-\mu^-(e^-\tau^
-,\mu^-\tau^-)
\label{3}
\ee
whose diagrams are drawn in Fig.[1].

Joining together the results obtained by Mohapatra and Swartz, we assume for
$g_{ee}$ in reaction (\ref{2}) the value corresponding to the upper limit
$g_{ee}^2=9.7\times 10^{-6}(M_H/1~GeV)^{2}~$
for $50<M_H<100~$GeV and the value
$g^2_{ee}=0.1$ for $M_H \ge 100~$GeV.
Moreover, in reaction (\ref{3}) we assume
Mohapatra's limit $g^2_{e\mu}=0.1$ that is the only bound for
$g_{e\mu}$. With these reasonable values of the
parameters, we find that it is possible to probe at HERA
doubly charged Higgs particles
with masses up to 150 GeV.

We computed the cross sections for the processes (\ref{2}) and (\ref{3}) using
the
Weizsacker-Williams method of the Equivalent Photons Approximation (EPA)
applied to the vertex $p\rightarrow p'\gamma^*$ (for a review of EPA
see \cite{Budnev}). In this approximation, the single $H^{--}$ production cross
section is given by
\be
\sigma_{ep}=\int N(\omega )\sigma_{e\gamma }(\omega )
{d\omega\over{\omega}}
\label{4}
\ee
where the quantity
\be
N(\omega )/\omega =\int dn(\omega ,q^2)dq^2
\label{5}
\ee
is the photon spectrum (see ref. \cite{Budnev}) and $\sigma_{e\gamma}(\omega )$
 is the cross section of the sub-process induced by the photon generated by
the proton current. The dependence of $N(\omega )$ on the photon frequency
$\omega$ is obtained by integrating the photon distribution $dn(\omega ,q^2
)$ over the squared momentum of the space-like photon, $q^2$. In the EPA,
 the $q^2$-dependence appears only in $dn(\omega ,q^2)$ whose
expression is determined by the structure of the $p\rightarrow p'\gamma^*$
hadronic vertex. Moreover, $\sigma_{e\gamma}$ is the cross section
for the absorbtion of a real unpolarized photon of frequency
$\omega$.

The electric and magnetic Sachs form factors of the proton
\cite{Sachs}, $G_E(q^2)$
and $G_M(q^2)$, have been included in the proton current to evaluate
$dn(\omega ,q^2)$.
We adopted the usual dipole expression for the Sachs form factors:
\be
G_E(q^2)={q_o^4\over{(q_o^2-q^2)^2}}~~~~~~~~~~~~~G_M(q^2)=\mu_pG_E(q^2)
\label{6}
\ee
where $\mu_p$ is the proton magnetic moment and with $q_o^2=0.71~GeV^2$.

We checked numerically that the necessary condition
for the validity of EPA
(the scalar and longitudinal photons contribution is much less than
the transverse one) is verified in the kinematic domain of our process.
More sophisticated checks that
confirm the applicability of EPA in our case are discussed in
ref.\cite{tesi}.

The amplitudes contributing to the $e^-\gamma\rightarrow e^+H^{--}$
interaction can be written as
\be
{\cal{M}}_a={g_{el}\over{2\sqrt{2}}}i\bar{u}(k')(1+a\gamma_5)
{(\hat{k}+\hat{q}+m_e)\over{(|k+q|^2-
m_e^2)}}\epsilon_{\alpha}(q)\gamma^{\alpha}u(k)
\label{7}
\ee
\be
{\cal{M}}_b={g_{el}\over{2\sqrt{2}}}i\bar{u}(k')\epsilon_{\alpha}(q)
\gamma^{\alpha}{(\hat{p}_H-\hat{k}+m_l)\over{(|p_H-k|^2-m_l^2)}}(1+a\gamma_5)
u(k)
\label{8}
\ee
\be
{\cal{M}}_c=2{g_{el}\over{2\sqrt{2}}}i\bar{u}(k')(1+a\gamma_5)u(k){(p_H+k-k')
_{\alpha}\epsilon^{\alpha}(q)\over{(|p_H-q|^2-M_H^2)}}
\label{9}
\ee
where $l$ is $e$ or $\mu$ and $a=+1$($-1$) for $H_R$ ($H_L$) production.

The analytical part of the calculation has been done by REDUCE and the results
for the total cross sections versus the $H$ boson's mass are plotted in
Fig.[2]. The resulting cross-sections
are independent of the sign of  $a$ and the same is true for the
decay's rate: $\Gamma(H_L \rightarrow ll')=\Gamma(H_R \rightarrow ll')$.
The differences between the two curves for $M_H< 100~$GeV are uniquely
due to the smaller values taken by the coupling costant $g_{ee}$ of the
process (\ref{2}), compared to $g_{e\mu}$. Both the cross sections,
$\sigma_{e\mu}$ and $\sigma_{ee}$, decrease rapidly when the value of $M_H$
increases (for $M_H=150~$GeV $\sigma =2.7\times 10^{-2}~$pb ). We stress that
the numerical integration is not completely straightforward. In fact, since
the energies involved in the process are much larger than the leptonic masses,
the electron (muon) propagator in ${\cal{M}}_b$ is very close to a pole.
We have accurately checked the
numerical convergence of the integration and we estimate that our results are
affected by a numerical error of about $10\%$.

In the final state, the $H^{--}$ longitudinal momentum has the same sign
of the incoming
proton's momentum,
since the relative velocity between the centre of mass
and HERA frames is always
greater than the $H^{--}$ velocity in the CMS. The angular
distributions of the positron and scattered proton are peaked along
the initial
proton direction. So, with high probability, the process is projected forward
on account of the particular kinematic of HERA.

In order to evaluate the final decay's leptons distribution at HERA, we
performed a Monte Carlo simulation of the $H^{--}$ decay. We adopted the value
1/3 for the branching ratio of both processes (\ref{2}) and (\ref{3}),
having assumed the $g_{ll'}$ values to be generation independent, i.e.
$g_{ee}\simeq g_{\mu\mu}\simeq g_{\tau\tau}$ and $g_{e\mu}\simeq
g_{e\tau}\simeq g_{\mu\tau}$ \cite{LEP}. The results are drawn in Fig.[3].
The distribution of one of the negative leptons in the final state
with respect to the angle $\theta$,
that is the angle between the direction of the incoming proton and the
outgoing decay's lepton, shifts at smaller values of $\theta$, as $M_H$
increases. Nevertheless this distribution
is still peaked at large angles up to $M_H=150~$GeV.
It is worth noting that the lepton pairs are produced with a large
opening angle, never lower than $15^o$ for $M_H\le 150~$GeV.

In order to estimate the number of observable events at HERA, we
assume \cite{H1} the range: $4^o\le\theta\le 176^o$ as the
angular acceptance of the HERA detectors and a
luminosity  $L=1.5\cdot 10^{31}~cm^{-2}s^{-1}$.

In Tab.[1], the number of
events/year as a function of $M_H$ are shown; we find that more than $92\%$ of
the lepton pairs overcome the angular cuts for $M_H>50~$GeV.
Increasing the value of $M_H$, the number of events rapidly decreases.
 Although the events are
few for $M_H\ge 120~$GeV, the detection of the $H^{--}$ may still be possible
even at large mass values.
In fact the signature of the event
is very particular because the proton goes into the pipe
and therefore the final state  consists of three leptons, two of them
(those coming from the $H^{--}$ decay)
with the same
sign and with large value of the invariant mass.

In conclusion, under reasonable assumption on the coupling costant values,
a single $H^{--}$ production may be observed at HERA through the $H$ decay in
a same-sign lepton pair up to $M_H$ $\simeq$ 150~GeV.

We strongly encourage an experimental investigation of the $H^{--}$ production
at HERA.

We thank M. Lusignoli for his suggestions, ideas and encouragement
throughout this work. We thank M. Iori, M. Mattioli and G. Penso for
useful discussions.

\vfill\eject

\vfill\eject

\newpage


\begin{table}
\label{tab-num}
\begin{tabular}{|c|c|c|c|c|c|c|c|c|}\hline\hline
\multicolumn{9}{|c|}{Number of events/year versus $M_H$ at HERA} \\
\hline\hline
           &       &   &      &      &      &      &       &        \\
$M_H$(GeV/$c^2$) & 50 &  60  &  70  &  80  &  90  &  100  &   120  &  150  \\
           &      & &  &     &      &      &       &        \\
\hline
                  &    &  &    &      &      &      &      &         \\
$N_{events}^{ee}(e^-p\rightarrow e^+pH^{--})$ & 95 & 75   &  60  &  48  &  38
&  31  &   14  &  4     \\
                  &  &      & &       &      &      &      &         \\
\hline
                  &   &   &   &   &      &      &      &        \\
$N_{events}^{e\mu}(e^-p\rightarrow\mu^+pH^{--})$   & 391  & 215   & 126  &
77  & 49   & 31  &  14  &  4  \\
                         &  & &  &  &      &      &      &        \\
\hline
\end{tabular}
\end{table}

\vspace{2 truecm}

\centerline{{\bf{TABLE and FIGURE CAPTIONS}}}
\vspace {1 truecm}
\noindent{\bf{TAB.1 :}}\quad Number of events/year versus $M_H$ at HERA for
the two reactions: $e^-p\rightarrow e^+pH^{--}~$ and $~e^-p\rightarrow \mu^+
pH^{--}$. The accuracy of the numerical calculation is of the order 10$\%$.

\vspace{.5 truecm}

\noindent{\bf{FIG.1 :}}\quad Feynman graphs for single $H^{--}$ production
at HERA.

\vspace {.5 truecm}

\noindent{\bf{FIG.2:}}\quad Plot of the total cross sections
at $\sqrt{s}=313~$GeV versus $M_H$: the dashed line refers
to the process $e^-p\rightarrow e^+pH^{--}$, the full line to the reaction
$e^-p\rightarrow\mu^+pH^{--}$.

\vspace {.5 truecm}

\noindent{\bf{FIG:3}}\quad The full, dashed and dot-dashed lines are the
normalized angular distribution of the decay's lepton at HERA for
$M_H$=70,100,150 GeV respectively.

%
%

\end{document}